\documentstyle[prd,aps,preprint,tighten,epsfig]{revtex}

\begin{document}

\draft

\title{Complete Parameter Space of Quark Mass Matrices \\
with Four Texture Zeros}
\author{{\bf Zhi-zhong Xing} ~ and ~ {\bf He Zhang}}
\address{Institute of High Energy Physics, 
Chinese Academy of Sciences, \\
P.O. Box 918 (4), Beijing 100039, China \\
({\it Electronic address: xingzz@mail.ihep.ac.cn}) }
\maketitle

\begin{abstract}
The full parameter space of Hermitian quark mass matrices 
with four texture zeros is explored by using current experimental
data. We find that all ten free parameters of the four-zero 
quark mass matrices can well be constrained. In particular, 
only one of the two phase parameters plays an important role 
in CP violation. The structural features of this specific pattern
of quark mass matrices are also discussed in detail.
\end{abstract}

\pacs{PACS number(s): 12.15.Ff, 12.10.Kt} 

\framebox{\Large\bf 1} ~
The texture of quark mass matrices, which can significantly impact on
the pattern of quark flavor mixing, is completely unknown in the standard 
electroweak model. A theory more fundamental than the standard model is
expected to allow us to determine the concrete structure of quark mass 
matrices, from which six quark masses, three flavor mixing angles and 
one CP-violating phase can fully be calculated. Attempts in this 
direction (e.g., those starting from supersymmetric grand unification 
theories and from superstring theories) are encouraging but have not 
proved to be very successful. Phenomenologically, a very common approach 
is to devise simple textures of quark mass matrices that can predict some 
self-consistent and experimentally-favored relations between quark masses 
and flavor mixing parameters \cite{Review}. The flavor symmetries hidden 
in such textures might finally provide us with useful hints about the 
underlying dynamics responsible for the generation of quark masses and 
the origin of CP violation. 

Without loss of generality, the quark mass matrices $M_{\rm u}$ 
(up-type) and $M_{\rm d}$ (down-type) can always be taken to be
Hermitian in the standard model or its extensions which have no 
flavor-changing right-handed currents \cite{F79}. Physics is invariant
under a common unitary transformation of Hermitian $M_{\rm u}$ and 
$M_{\rm d}$ (i.e., $M_{\rm u,d} \rightarrow S M_{\rm u,d} S^\dagger$ 
with $S$ being an arbitrary unitary matrix). This freedom allows a
further arrangement of the structures of quark mass matrices, such that 
\begin{equation}
M_{\rm u} \; = \; \left ( \matrix{
{\bf 0} & C_{\rm u}     & {\bf 0}  \cr
C^*_{\rm u}     & \tilde{B}_{\rm u}       & B_{\rm u} \cr
{\bf 0} & B^*_{\rm u}   & A_{\rm u} \cr} \right ) \; 
\end{equation}
and
\begin{equation}
M_{\rm d} \; = \; \left ( \matrix{
D_{\rm d} & C_{\rm d}   & {\bf 0} \cr
C^*_{\rm d}     & \tilde{B}_{\rm d}       & B_{\rm d} \cr
{\bf 0} & B^*_{\rm d}   & A_{\rm d} \cr} \right ) \; 
\end{equation}
or
\begin{equation}
M'_{\rm d} \; = \; \left ( \matrix{
{\bf 0} & C_{\rm d}     & D'_{\rm d} \cr
C^*_{\rm d}     & \tilde{B}_{\rm d}       & B_{\rm d} \cr
{D'}^*_{\rm d}      & B^*_{\rm d}   & A_{\rm d} \cr} \right ) \; 
\end{equation}
hold \cite{FX99}. We see that $M_{\rm u}$ has two texture zeros and 
$M_{\rm d}$ or $M'_{\rm d}$ has one texture zero \cite{Note}.
Because the texture zeros of quark mass matrices in Eqs. (1)--(3)
result from some proper transformations of the flavor basis 
under which the gauge currents keep diagonal and real, there 
is no loss of any physical content for quark masses and flavor
mixing. But it is impossible to further obtain 
$D_{\rm d} = 0$ for $M_{\rm d}$ or $D'_{\rm d} =0$ for 
$M'_{\rm d}$ via a new physics-irrelevant transformation of 
the flavor basis \cite{FX99}. 
In other words, $D_{\rm d} = D'_{\rm d} = 0$ can only be a
physical assumption. This assumption leads to the well-known
four-zero texture of Hermitian quark mass matrices, which has the 
up-down parallelism and respects the chiral evolution of quark 
masses \cite{F87}.

Although a lot of interest has been paid to the four-zero texture
of Hermitian quark mass matrices \cite{4zero,FX03}, 
a complete analysis of its parameter space has been lacking. 
This unsatisfactory situation is partly due to the fact that many 
authors prefer to make instructive analytical approximations in 
analyzing the consequences of $M_{\rm u}$
and $M_{\rm d}$ on flavor mixing and CP violation. Some non-trivial
parts of the whole parameter space of $M_{\rm u,d}$ were 
unfortunately missed or ignored in such analytical approximations 
for a long time, as pointed out in Ref. \cite{FX03}. 

The purpose of this short note is to make use of current experimental
data to explore the complete parameter space of the four-zero
quark mass matrices $M_{\rm u}$ and $M_{\rm d}$. We find that 
the ten free parameters of $M_{\rm u}$ and $M_{\rm d}$ can well be 
constrained. In particular, only one of the two phase parameters 
plays an important role in CP violation. We shall also discuss the 
structural features of $M_{\rm u}$ and $M_{\rm d}$ in detail.

\framebox{\Large\bf 2} ~
Let us concentrate on the four-zero texture of Hermitian quark mass 
matrices given in Eqs. (1) and (2) with $D_{\rm d}=0$. 
The observed hierarchy of quark masses
($m_u \ll m_c \ll m_t$ and $m_d \ll m_s \ll m_b$) implies that
$|A_{\rm q}| > |\tilde{B}_{\rm q}|, |B_{\rm q}| > |C_{\rm q}|$ 
(for q = u or d) should in general hold \cite{F87}. Note that
$M_{\rm q}$ can be decomposed into
$M_{\rm q} = P^\dagger_{\rm q} \overline{M}_{\rm q} P_{\rm q}$, where
\begin{equation}
\overline{M}_{\rm q} \; = \; \left ( \matrix{
{\bf 0} & |C_{\rm q}|   & {\bf 0} \cr
|C_{\rm q}|     & \tilde{B}_{\rm q}     & |B_{\rm q}| \cr
{\bf 0} & |B_{\rm q}|   & A_{\rm q} \cr} \right ) \; 
\end{equation}
and $P_{\rm q} = {\rm Diag} \{1, \exp(i\phi^{~}_{C_{\rm q}}),
\exp(i\phi^{~}_{B_{\rm q}} + i\phi^{~}_{C_{\rm q}}) \}$ with 
$\phi^{~}_{B_{\rm q}} \equiv \arg (B_{\rm q})$ and 
$\phi^{~}_{C_{\rm q}} \equiv \arg (C_{\rm q})$. 
For simplicity, we neglect the subscript ``q'' in the following,
whenever it is unnecessary to distinguish between the up and down quark 
sectors. The real symmetric mass matrix $\overline{M}$ can be diagonalized
by use of the orthogonal transformation
\begin{equation}
O^T \overline{M} O \; = \; \left ( \matrix{
\lambda_1 & 0 & 0 \cr
0 & \lambda_2 & 0 \cr
0 & 0 & \lambda_3 \cr} \right ) \; ,
\end{equation}
where $\lambda_i$ (for $i=1,2,3$) are quark mass eigenvalues.
Without loss of generality, we take $\lambda_3 >0$ and $A >0$. 
Then ${\rm Det} (\overline{M}) = -A|C|^2 < 0$ implies that 
$\lambda_1 \lambda_2 < 0$ is required. It is easy to find that 
$\tilde{B}$, $|B|$ and $|C|$ can be expressed in terms of 
$\lambda_i$ and $A$ as 
\begin{eqnarray}
\tilde{B} & = & \lambda_1 + \lambda_2 + \lambda_3 - A \; , 
\nonumber \\
|B| & = & \sqrt{\frac{(A -\lambda_1) (A -\lambda_2) 
(\lambda_3 -A)}{A}} \;\; ,
\nonumber \\
|C| & = & \sqrt{\frac{-\lambda_1 \lambda_2 \lambda_3}{A}} \;\; . 
\end{eqnarray}
The exact expression of $O$ turns out to be \cite{FX03}
\small
\begin{equation}
O \; =\; \left ( \matrix{ \cr
\displaystyle
\sqrt{\frac{\lambda_2 \lambda_3 (A-\lambda_1)}{A (\lambda_2 - \lambda_1)
(\lambda_3 - \lambda_1)}}
& \displaystyle
\eta  \sqrt{\frac{\lambda_1 \lambda_3 (\lambda_2 -A)}
{A (\lambda_2 - \lambda_1) (\lambda_3 - \lambda_2)}}
& \displaystyle
\sqrt{\frac{\lambda_1 \lambda_2 (A -\lambda_3)}
{A (\lambda_3 - \lambda_1) (\lambda_3 - \lambda_2)}} \cr\cr\cr
\displaystyle 
- \eta \sqrt{\frac{\lambda_1 (\lambda_1 -A)}{(\lambda_2 - \lambda_1)
(\lambda_3 - \lambda_1)}} 
& \displaystyle
\sqrt{\frac{\lambda_2 (A-\lambda_2)}
{(\lambda_2 - \lambda_1) (\lambda_3 - \lambda_2)}}
& \displaystyle
\sqrt{\frac{\lambda_3 (\lambda_3 -A)}
{(\lambda_3 - \lambda_1) (\lambda_3 - \lambda_2)}} \cr\cr\cr
\displaystyle
\eta \sqrt{\frac{\lambda_1 (A -\lambda_2) (A -\lambda_3)}
{A (\lambda_2 - \lambda_1) (\lambda_3 - \lambda_1)}}
& \displaystyle
- \sqrt{\frac{\lambda_2 (A-\lambda_1) (\lambda_3 -A)}
{A (\lambda_2 - \lambda_1) (\lambda_3 - \lambda_2)}}
& \displaystyle
\sqrt{\frac{\lambda_3 (A -\lambda_1) (A -\lambda_2)}
{A (\lambda_3 - \lambda_1) (\lambda_3 - \lambda_2)}} \cr\cr}
\right ) \; ,
\end{equation}
\normalsize
in which $\eta \equiv \lambda_2/m_2 = +1$ or $-1$ corresponding to
the possibility $(\lambda_1, \lambda_2) = (-m_1, +m_2)$ or
$(+m_1, -m_2)$. The Cabibbo-Kobayashi-Maskawa (CKM) flavor mixing 
matrix \cite{CKM}, which measures the non-trivial mismatch between
diagonalizations of $M_{\rm u}$ and $M_{\rm d}$, is given by
$V \equiv O^T_{\rm u} (P_{\rm u} P^\dagger_{\rm d}) O_{\rm d}$.
Explicitly, we have 
\begin{equation}
V_{i\alpha} \; =\; O^{\rm u}_{1i} O^{\rm d}_{1\alpha} +
O^{\rm u}_{2i} O^{\rm d}_{2\alpha} e^{i\phi_1} +
O^{\rm u}_{3i} O^{\rm d}_{3\alpha} e^{i(\phi_1 + \phi_2)} \; ,
\end{equation}
where the subscripts $i$ and $\alpha$ run respectively over 
$(u,c,t)$ and $(d,s,b)$, and two phases are defined as
$\phi_1 \equiv \phi^{~}_{C_{\rm u}} - \phi^{~}_{C_{\rm d}}$ and
$\phi_2 \equiv \phi^{~}_{B_{\rm u}} - \phi^{~}_{B_{\rm d}}$.

It is well known that nine elements of the CKM matrix $V$ have 
six orthogonal relations, corresponding to six triangles in the 
complex plane \cite{Review}. Among them, the unitarity triangle 
defined by $V^*_{ub}V_{ud} + V^*_{cb}V_{cd} + V^*_{tb}V_{td} =0$
is of particular interest for the study of CP violation at
$B$-meson factories \cite{PDG}. Three inner angles of this triangle
are commonly denoted as 
\begin{eqnarray}
\alpha & = & \arg \left ( - \frac{V^*_{tb}V_{td}}{V^*_{ub}V_{ud}}
\right ) \;\; , \nonumber \\
\beta & = & \arg \left ( -\frac{V^*_{cb}V_{cd}}{V^*_{tb}V_{td}}
\right ) \;\; , \nonumber \\
\gamma & = & \arg \left ( -\frac{V^*_{ub}V_{ud}}{V^*_{cb}V_{cd}}
\right ) \;\; .
\end{eqnarray}
So far the angle $\beta$ has unambiguously been measured from
the CP-violating asymmetry in $B^0_d$ vs 
$\bar{B}^0_d\rightarrow J/\psi K_{\rm S}$ decays \cite{Browder}. 
The angles $\alpha$ and $\gamma$ are expected to be detected at 
KEK and SLAC $B$-meson factories in the near future. Given the 
four-zero texture of quark mass matrices, these three angles depend 
on the CP-violating phases $\phi_1$ and $\phi_2$. We shall examine
the explicit dependence of $(\alpha, \beta, \gamma)$ on
$(\phi_1, \phi_2)$ in the following.

\framebox{\Large\bf 3} ~
Now we explore the whole parameter space of $M_{\rm u}$ and
$M_{\rm d}$ with the help of current experimental data. There
are totally ten free parameters associated with $M_{\rm u,d}$:
$A_{\rm u,d}$, $|B_{\rm u,d}|$, $\tilde{B}_{\rm u,d}$,
$|C_{\rm u,d}|$ and $\phi_{1,2}$. In comparison, there are
also ten observables which can be derived from the four-zero
texture of quark mass matrices: six quark masses 
($m_u, m_c, m_t$ and $m_d, m_s, m_b$) and four independent
parameters of quark flavor mixing (typically, $|V_{us}|$, 
$|V_{cb}|$, $|V_{ub}/V_{cb}|$ and $\sin 2\beta$). Thus there is 
no problem to determine the complete parameter space of $M_{\rm u,d}$.

(1) The first step of our numerical calculations is to find out the
allowed ranges of $A_{\rm u}/m_t$, $A_{\rm d}/m_b$, $\phi_1$ and 
$\phi_2$ by using Eqs. (6)-(9). For this purpose, we adopt the
following reasonable and generous values of quark mass ratios at the 
electroweak scale $\mu = M_Z$ \cite{Review,PDG}:
\begin{eqnarray}
\frac{m_c}{m_u} & = & 270 - 350 \; , ~~~~~
\frac{m_s}{m_d} \; = \; 17 - 25 \; ;
\nonumber \\
\frac{m_t}{m_c} & = & 260 - 320 \; , ~~~~~
\frac{m_b}{m_s} \; = \; 35 - 45 \; .
\end{eqnarray}
The predictions of $M_{\rm u}$ and $M_{\rm d}$ for the CKM
matrix elements are required to agree with current experimental 
data \cite{Browder,Buras}: 
\begin{eqnarray}
|V_{us}| & = & 0.2240 \pm 0.0036 \; , ~~~~~~~~~
\left | \frac{V_{ub}}{V_{cb}} \right | \; =\; 0.086 \pm 0.008 \; ,
\nonumber \\
|V_{cb}| & = & (41.5 \pm 0.8) \times 10^{-3} \; , ~~~~~
\sin 2\beta \; =\; 0.736 \pm 0.049 \; .
\end{eqnarray}
Note that the results of $V$ may involve a four-fold ambiguity
arising from four possible values of $(\eta_{\rm u}, \eta_{\rm d})$
in $O_{\rm u}$ and $O_{\rm d}$, as one can see from Eq. (7). To be 
specific, we first choose $\eta_{\rm u} = \eta_{\rm d} =+1$ in
our numerical analysis and then discuss the other three possibilities.

The numerical results for $A_{\rm u}/m_t$ vs $A_{\rm d}/m_b$
and $\phi_1$ vs $\phi_2$ are illustrated in Fig. 1. We see that the
most favorable values of these four quantities are 
$A_{\rm u}/m_t \sim 0.94$, $A_{\rm d}/m_b \sim 0.94$,
$\phi_1 \sim 0.5\pi$ and $\phi_2 \sim 1.96\pi$. Fig. 1(a) confirms
that the (3,3) elements (i.e., $A_{\rm u} \sim m_t$ and 
$A_{\rm d} \sim m_b$) are the dominant matrix elements in $M_{\rm u}$ 
and $M_{\rm d}$. The strong constraint on $\phi_1$ comes from  
the experimental data on $|V_{us}|$ and $\sin 2\beta$; while the
tight restriction on $\phi_2$ results from current data on 
$|V_{cb}|$. Because of $\sin \phi_1 \gg |\sin \phi_2|$ as shown in
Fig. 1(b), the strength of CP violation in the CKM matrix is mainly 
governed by $\phi_1$. In many analytical approximations, 
$\phi_1 \approx 0.5\pi$ and $\phi_2 =0$ have typically been 
taken \cite{4zero,FX03}.
 
We find that both $A_{\rm u}/m_t$ and $A_{\rm d}/m_b$ are 
insensitive to the signs of $\eta_{\rm u}$ and $\eta_{\rm d}$.
In other words, the allowed ranges of $A_{\rm u}/m_t$ and 
$A_{\rm d}/m_b$ are essentially the same in 
$(\eta_{\rm u}, \eta_{\rm d}) = (\pm 1, \pm1)$ and 
$(\pm 1, \mp1)$ cases. While $\phi_1$ is sensitive
to the signs of $\eta_{\rm u}$ and $\eta_{\rm d}$, 
$\phi_2$ is not. To be explicit, we have
\begin{eqnarray}
(\eta_{\rm u}, \eta_{\rm d}) & = & (+1, +1): ~~~~~
\phi_1 \sim 0.5\pi \; , ~~~ \phi_2 \lesssim 2\pi \; ,
\nonumber \\
(\eta_{\rm u}, \eta_{\rm d}) & = & (+1, -1): ~~~~~
\phi_1 \sim 1.5\pi \; , ~~~ \phi_2 \gtrsim 0 \; ,
\nonumber \\
(\eta_{\rm u}, \eta_{\rm d}) & = & (-1, +1): ~~~~~
\phi_1 \sim 1.5\pi \; , ~~~ \phi_2 \lesssim 2\pi \; ,
\nonumber \\
(\eta_{\rm u}, \eta_{\rm d}) & = & (-1, -1): ~~~~~
\phi_1 \sim 0.5\pi \; , ~~~ \phi_2 \gtrsim 0 \; .
\end{eqnarray}
The dependence of $\phi_1$ on $\eta_{\rm u}$ and $\eta_{\rm d}$
can easily be understood. Indeed, 
$\tan\beta \propto \eta_{\rm u}\eta_{\rm d} \sin\phi_1$ holds
in the leading-order analytical approximation with $|\sin\phi_2| \ll 1$.
Thus the positiveness of $\tan\beta$ requires that $\sin\phi_1$ and
$\eta_{\rm u}\eta_{\rm d}$ have the same sign.

(2) The second step of our numerical analysis is to determine 
the relative magnitudes of four non-zero matrix elements of 
$M_{\rm u,d}$ by using Eq. (6) and the results for 
$A_{\rm u}/m_t$ and $A_{\rm d}/m_b$. The numerical results for
$|B_{\rm u}|/A_{\rm u}$ vs $|B_{\rm d}|/A_{\rm d}$,
$\tilde{B}_{\rm u}/|B_{\rm u}|$ vs $\tilde{B}_{\rm d}/|B_{\rm d}|$ 
and $|C_{\rm u}|/\tilde{B}_{\rm u}$ vs $|C_{\rm d}|/\tilde{B}_{\rm d}$ 
are shown in Fig. 2. One can see that the most favorable values of
these six quantities are
$|B_{\rm u}|/A_{\rm u} \sim 0.25$, 
$\tilde{B}_{\rm u}/|B_{\rm u}| \sim 0.3$,
$|C_{\rm u}|/\tilde{B}_{\rm u} \sim 0.003$ and
$|B_{\rm d}|/A_{\rm d} \sim 0.25$, 
$\tilde{B}_{\rm d}/|B_{\rm d}| \sim 0.4$,
$|C_{\rm d}|/\tilde{B}_{\rm d} \sim 0.06$. 
A remarkable feature of our typical results is that $A_{\rm q}$, 
$|B_{\rm q}|$ and $\tilde{B}_{\rm q}$ (for q = u or d) roughly 
satisfy a geometric relation: 
$|B_{\rm q}|^2 \sim A_{\rm q} \tilde{B}_{\rm q}$ \cite{FX03}.
In addition, $|B_{\rm u}| \gg m_c$ and $|B_{\rm d}| \gg m_s$ hold.
While a very strong hierarchy exists between (1,2) and (2,2) elements 
of $M_{\rm u,d}$, there is only a weak hierarchy among (2,2), (2,3) 
and (3,3) elements of $M_{\rm u,d}$. Such a structural property of
quark mass matrices must be taken into account in model building.

To be more explicit, let us illustrate the texture of 
$\overline{M}_{\rm u,d}$ by choosing $m_c/m_u = 320$, $m_t/m_c = 290$,
$m_s/m_d =21$ and $m_b/m_s = 40$. We obtain
\begin{eqnarray}
\overline{M}_{\rm u} & \approx & A_{\rm u} \left ( \matrix{
{\bf 0} & 0.0002 & {\bf 0} \cr
0.0002 & 0.067 & 0.24 \cr
{\bf 0} & 0.24 & {\bf 1} \cr} \right ) \sim \; 
A_{\rm u} \left ( \matrix{
{\bf 0} & \varepsilon^6 & {\bf 0} \cr
\varepsilon^6 & \varepsilon^2 & \varepsilon \cr
{\bf 0} & \varepsilon & {\bf 1} \cr} \right ) \; ,
\nonumber \\
\overline{M}_{\rm d} & \approx & A_{\rm d} \left ( \matrix{
{\bf 0} & 0.0059 & {\bf 0} \cr
0.0059 & 0.089 & 0.24 \cr
{\bf 0} & 0.24 & {\bf 1} \cr} \right ) \sim \;
A_{\rm d} \left ( \matrix{
{\bf 0} & \varepsilon^4 & {\bf 0} \cr
\varepsilon^4 & \varepsilon^2 & \varepsilon \cr
{\bf 0} & \varepsilon & {\bf 1} \cr} \right ) \; ,
\end{eqnarray}
where $\varepsilon \approx 0.24$ in this special case \cite{Note2}.
Such a four-zero pattern of quark mass matrices depends on a
small expansion parameter and is quite suggestive for model building.
For example, one may speculate that $\overline{M}_{\rm u}$ and 
$\overline{M}_{\rm d}$ in Eq. (13)
could naturally result from a string-inspired model of
quark mass generation \cite{Ibanez} or from a horizontal U(1) family 
symmetry and its perturbative breaking \cite{Flavor}.

Note that the numerical results in Fig. 2 and Eq. (13) have been 
obtained by taking $\eta_{\rm u} = \eta_{\rm d} =+1$. A careful
analysis shows that $\tilde{B}_{\rm q}$ (for q = u or d) is 
sensitive to the sign of $\eta_{\rm q}$, but $|B_{\rm q}|$ and 
$|C_{\rm q}|$ are not. In view of Eq. (6), we find that the sign
of $\eta = \lambda_2/m_2$ may significantly affect the size
of $\tilde{B}$ if its $\lambda_2$ and $\lambda_3 -A$ terms are 
comparable in magnitude. In contrast, the dependence of $|B|$ on
$\eta$ is negligible due to $A \gg m_2$; and $|C|$ is completely
independent of the sign of $\eta$. 

(3) The final step of our numerical analysis is to examine the
outputs of three CP-violating angles $(\alpha, \beta, \gamma)$ and
the ratio $|V_{ub}/V_{cb}|$ constrained by the four-zero texture
of quark mass matrices. We plot the result for $|V_{ub}/V_{cb}|$
vs $\sin 2\beta$ in Fig. 3(a) and that for $\alpha$ vs $\gamma$ in
Fig. 3(b). The correlation between $\alpha$ and $\gamma$ is quite
obvious, as a result of $\alpha + \beta + \gamma = \pi$. Typically,
$\alpha \sim 0.5\pi$ holds. The possibility $\alpha \approx \gamma$,
implying that the unitarity triangle is approximately an 
isoceless triangle \cite{FH}, is also allowed by current data 
and quark mass matrices with four texture zeros. Note that the
size of $\sin 2\beta$ increases with $|V_{ub}/V_{cb}|$. This feature
can easily be understood: in the unitarity triangle with three
sides rescaled by $|V_{cb}|$, the inner angle $\beta$ corresponds 
to the side proportional to $|V_{ub}/V_{cb}|$.

Finally we mention that the outputs of $|V_{ub}/V_{cb}|$ and
$(\alpha, \beta, \gamma)$ are completely insensitive to the sign
ambiguity of $\eta_{\rm u}$ and $\eta_{\rm d}$.

\framebox{\Large\bf 4} ~
In summary, we have analyzed the complete parameter space of 
Hermitian quark mass matrices with four texure zeros by using
current experimental data. It is clear that the four-zero 
pattern of quark mass matrices can survive current experimental 
tests and its parameter space gets well constrained. We find
that only one of the two phase parameters plays a crucial role
in CP violation. The (2,2), (2,3) and (3,3) elements of the
up- or down-type quark mass matrix have a relatively weak 
hierarchy, although their magnitudes are considerably larger
than the magnitude of the (1,2) element. Such a structural 
feature of the four-zero quark mass matrices might serve as 
a useful starting point of view for model building.

We remark that the phenomenological consequences of quark mass 
matrices depend both on the number of their texture zeros and 
on the hierarchy of their non-vanishing entries. The former 
are in general not preserved to all orders or at any energy scales 
in the unspecified interactions which generate quark masses 
and flavor mixing \cite{X03}. But an experimentally-favored 
texture of quark mass matrices at low energy scales (such as the
one under discussion) is possible to shed some light on the 
underlying flavor symmetry and its breaking mechanism responsible
for fermion mass generation and CP violation at high energy scales.

\vspace{0.4cm}

This work was supported in part by National Natural Science 
Foundation of China.

\vspace{0.4cm}

\hspace{-0.8cm} {\bf Note added}: While our paper was being 
completed, we received a preprint by Zhou \cite{Zhou}, in which 
all possible four-zero textures of quark mass matrices are 
classified and computed. The analyses, results and discussions 
in these two papers have little overlap.

\begin{figure}[t]
\vspace{-1cm}
\epsfig{file=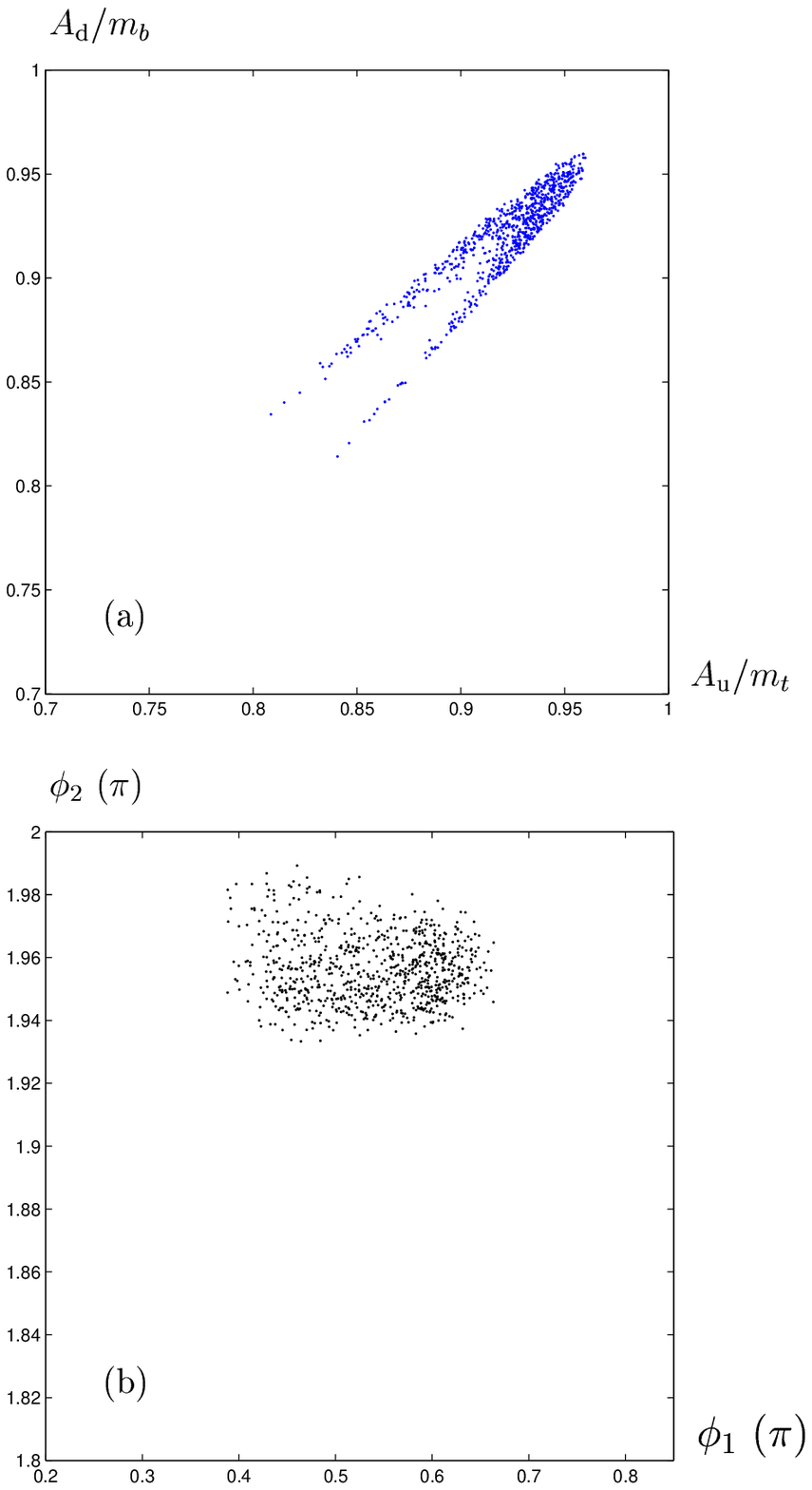,bbllx=2.5cm,bblly=12cm,bburx=17.5cm,bbury=30cm,%
width=13cm,height=15cm,angle=0,clip=90}
\vspace{4.4cm}
\caption{The allowed ranges of $A_{\rm u}/m_t$ vs 
$A_{\rm d}/m_b$ and $\phi_1$ vs $\phi_2$ for the four-zero
texture of quark mass matrices with 
$\eta_{\rm u} = \eta_{\rm d} = +1$.}
\end{figure}

\begin{figure}[t]
\vspace{-5.5cm}
\epsfig{file=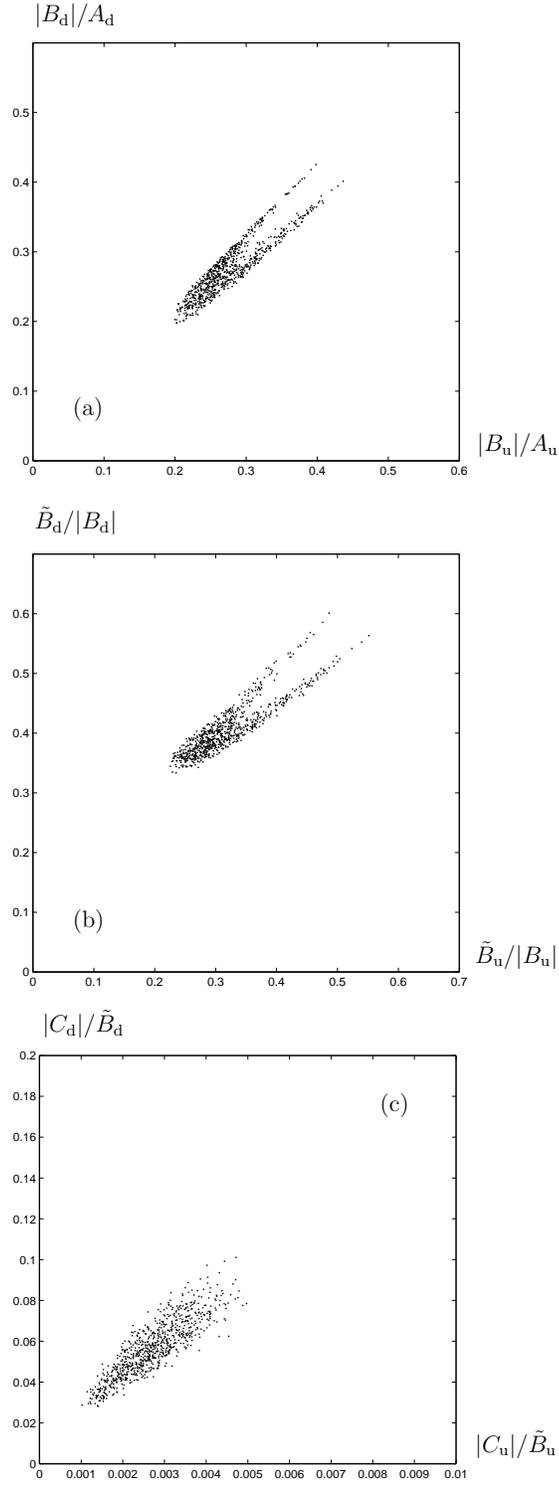,bbllx=1.5cm,bblly=10cm,bburx=18.5cm,bbury=30cm,%
width=13cm,height=15cm,angle=0,clip=90}
\vspace{6.4cm}
\caption{The allowed ranges of $|B_{\rm u}|/A_{\rm u}$ vs 
$|B_{\rm d}|/A_{\rm d}$, $\tilde{B}_{\rm u}/|B_{\rm u}|$ vs
$\tilde{B}_{\rm d}/|B_{\rm d}|$ and $|C_{\rm u}|/\tilde{B}_{\rm u}$ vs 
$|C_{\rm d}|/\tilde{B}_{\rm d}$ for the four-zero texture of quark 
mass matrices with $\eta_{\rm u} = \eta_{\rm d} = +1$.}
\end{figure}

\begin{figure}[t]
\vspace{-3cm}
\epsfig{file=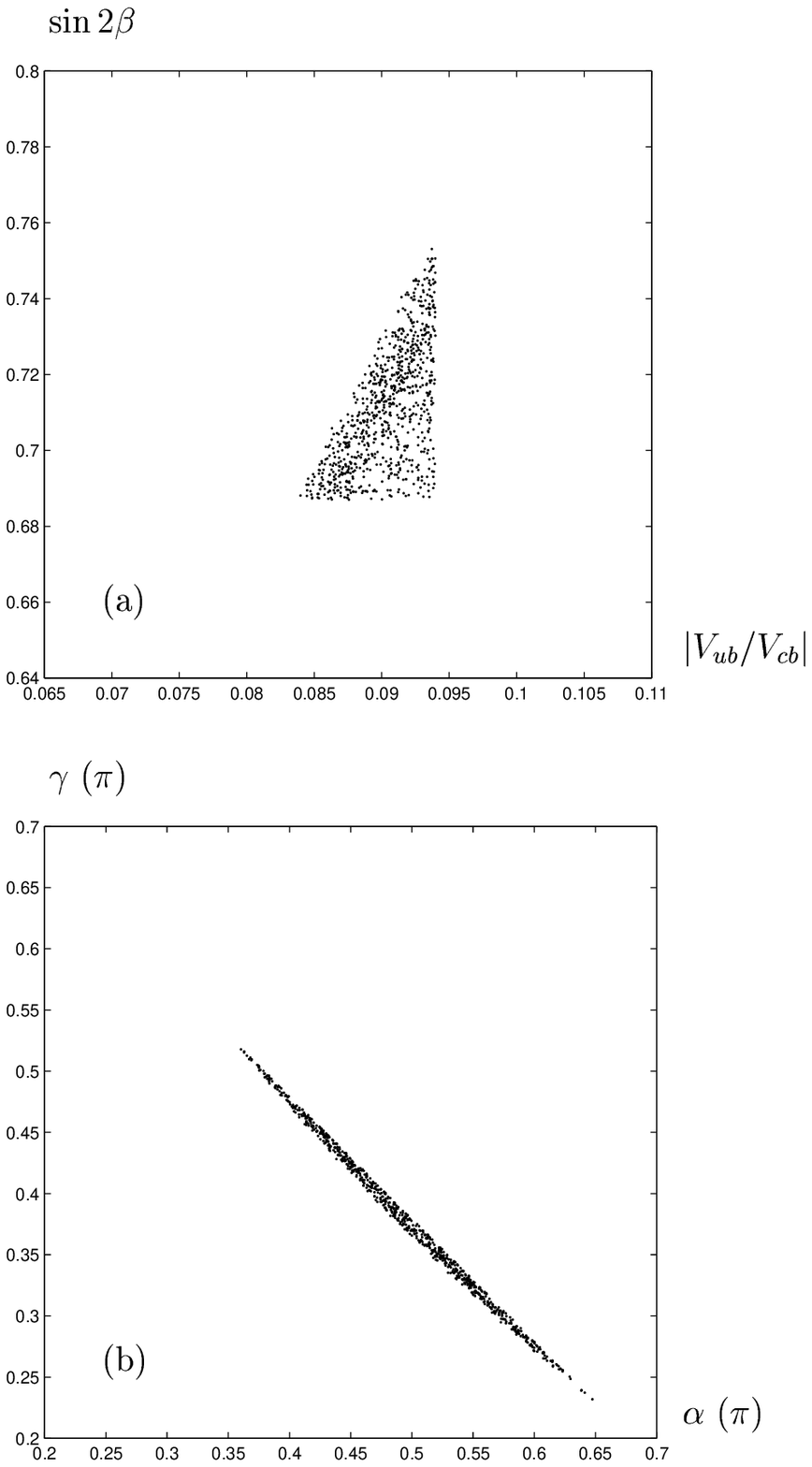,bbllx=2.5cm,bblly=12cm,bburx=17.5cm,bbury=30cm,%
width=13cm,height=15cm,angle=0,clip=90}
\vspace{4.4cm}
\caption{The allowed ranges of $|V_{ub}/V_{cb}|$ vs 
$\sin 2\beta$ and $\alpha$ vs $\gamma$ for the four-zero
texture of quark mass matrices with 
$\eta_{\rm u} = \eta_{\rm d} = +1$.}
\end{figure}

\begin{thebibliography}{99}
\bibitem{Note} Because $M_{\rm u}$ and $M_{\rm d}$ (or $M'_{\rm d}$)
are Hermitian, a pair of off-diagonal texture zeros in each
mass matrix have been counted as one zero.
\end{thebibliography}
\end{document}